\documentclass[lletterpaper,etter]{jpsj3}
\usepackage{txfonts}

\usepackage{epsfig}
\usepackage{graphicx}
\usepackage{dcolumn}
\usepackage{bm}
\usepackage{color} 

\title{Anomalous Superconducting-Gap Structure of Slightly Overdoped  Ba(Fe$_{1-x}$Co$_{x}$)$_{2}$As$_{2}$}

\author{Tetsuya Hajiri$^{1,2}$\thanks{Email: hajiri.tetsuya@f.mbox.nagoya-u.ac.jp}, Takahiro Ito$^{1,3}$, Masaharu Matsunami$^{2,4}$, Byeong Hun Min$^5$, Yong Seung Kwon$^{5}$, and Shin-ichi Kimura$^{2,4}$\thanks{Present address: Graduate School of Frontier Biosciences, Osaka University, Suita 565-0871, Japan}}
\inst{$^1$Graduate School of Engineering, Nagoya University, Nagoya 464-8603, Japan \\
$^2$UVSOR Facility, Institute for Molecular Science, Okazaki 444-8585, Japan \\
$^3$Nagoya University Synchrotron Radiation Research Center, Nagoya University, Nagoya, 464-8603, Japan \\
$^4$School of Physical Sciences, The Graduate University for Advanced Studies (SOKENDAI), Okazaki 444-8585, Japan \\
$^5$Department of Emerging Materials Science, DGIST, Daegu 711-873, Republic of Korea} %

\abst{
We observed the anisotropic superconducting-gap (SC-gap) structure of a slightly overdoped superconductor, Ba(Fe$_{1-x}$Co$_{x}$)$_{2}$As$_{2}$ ($x=0.1$), using three-dimensional (3D) angle-resolved photoemission spectroscopy.
Two hole Fermi surfaces (FSs) observed at the Brillouin zone center and an inner electron FS at the zone corner showed a nearly isotropic SC gap in 3D momentum space. 
However, the outer electron FS showed an anisotropic SC gap with nodes or gap minima around the M and A points.
The different anisotropies obtained the SC gap between the outer and inner electron FSs cannot be expected from all theoretical predictions with spin fluctuation, orbital fluctuation, and both competition.
Our results provide a new insight into the SC mechanisms of iron pnictide superconductors. 
}

\begin{document}
\maketitle

Iron pnictide superconductors~\cite{ref1} have the second-highest superconducting transition temperature ($T_c$) and a wide variety of superconducting (SC) states, particularly as SC-gap structure owing to multiple Fermi surfaces (FSs) and orbital characters~\cite{ref2,ref3}.
Focusing on the SC state of the 122-type iron pnictides, the Ba$_{1-x}$K$_{x}$Fe$_{2}$As$_{2}$ system shows a fully opened SC gap and a nodal SC gap around optimally doped~\cite{ref4} and a hole-end KFe$_2$As$_2$~\cite{ref5}, respectively, observed by angle-resolved photoemission spectroscopy (ARPES)~\cite{ref6,ref7,ref8,ref9}.
On the other hand, the BaFe$_{2}$(As$_{1-x}$P$_{x}$)$_{2}$ system has nodes in the whole SC region~\cite{ref10,ref11}.
There are different ARPES results for this system: one reports nodes at the hole FS of the $d_{z^2}$ orbital around the Z point~\cite{ref12}, and the other around the longer axis of the ellipsoid-shaped electron FS despite the lack of node in the hole FSs~\cite{ref13}.
The former can be explained by spin fluctuation alone~\cite{ref14}, while the latter requires an additional orbital fluctuation~\cite{ref15}.

The Ba(Fe$_{1-x}$Co$_{x}$)$_{2}$As$_{2}$ system shows an antiferromagnetic spin fluctuation in the whole of SC region~\cite{ref16}.
The antiferromagnetic scattering vector {\boldmath $Q$}$_{\rm AF}$ is from a hole FS to an electron FS~\cite{ref17}, which is similar to those in other iron pnictides.
Hence, spin fluctuation that induces a fully opened $s_{\pm}$-wave~\cite{ref18,ref19} in this system has been proposed.
On the other hand, both the softening of the $C_{66}$ mode~\cite{ref20} and the orbital-polarized electronic structure~\cite{ref21} suggest the presence of orbital fluctuation.
It is therefore important to elucidate the effects of spin and orbital fluctuations on the electronic structure.
With respect to the SC state, a fully opened SC gap has been observed in the underdoped region of $x\leq0.075$~\cite{ref22,ref23}, although some experiments have shown evidence of nodal behavior~\cite{ref24,ref25}.
In the overdoped region of $x>0.075$, on the other hand, a nodal SC gap with either nodes or gap minima on at least one FS has been suggested~\cite{ref23,ref24,ref25,ref26}.
To clarify the SC state of this unique system, it is necessary to elucidate the position of nodes in the overdoped region.
In this system, only one ARPES study of the SC-gap structure of optimally doped Ba(Fe$_{1-x}$Co$_{x}$)$_{2}$As$_{2}$ has been reported~\cite{ref27}, while no ARPES studies of the SC gap or nodes of overdoped  Ba(Fe$_{1-x}$Co$_{x}$)$_{2}$As$_{2}$ have been reported.
Since a three-dimensional (3D) SC-gap structure can be expected, it is important for ARPES in 3D momentum space using tunable incident photons.

In this Letter, we report our investigation of the 3D SC-gap structure of slightly overdoped Ba(Fe$_{1-x}$Co$_{x}$)$_{2}$As$_{2}$ $(x=0.1)$, using synchrotron radiation ARPES.
The two hole FSs obtained at the zone center and the inner electron FS obtained at the zone corner were found to have nearly isotropic SC gaps in 3D momentum space.
In the outer electron FS, however, we observed an anisotropic SC gap with nodes or gap minima.
The different anisotropies obtained the SC gap between the outer and inner electron FSs is not consistent with previous theories.

The 3D-ARPES experiments were performed on $in$-$situ$ cleaved single crystals of slightly overdoped Ba(Fe$_{1-x}$Co$_{x}$)$_{2}$As$_{2}$ $(x=0.1)$ with $T_c\sim22.5$~K.
The ARPES measurements were carried out at the “SAMRAI” end-station of the undulator beamline BL7U of UVSOR-I\hspace{-.1em}I\hspace{-.1em}I  at the Institute for Molecular Science, using an MBS A-1 analyzer~\cite{ref28}. 
The energy resolution was about 5 meV for the SC-gap measurement and in-plane FS mapping, and about 10~meV for the band-dispersion mapping and out-of-plane FS mapping.
The angular resolution was about  $0.17^{\circ}$. 
The Fermi level ($E_{\rm F}$) was calibrated using an evaporated gold film.

\begin{figure}[t]
\begin{center}
\includegraphics[width=8cm,clip]{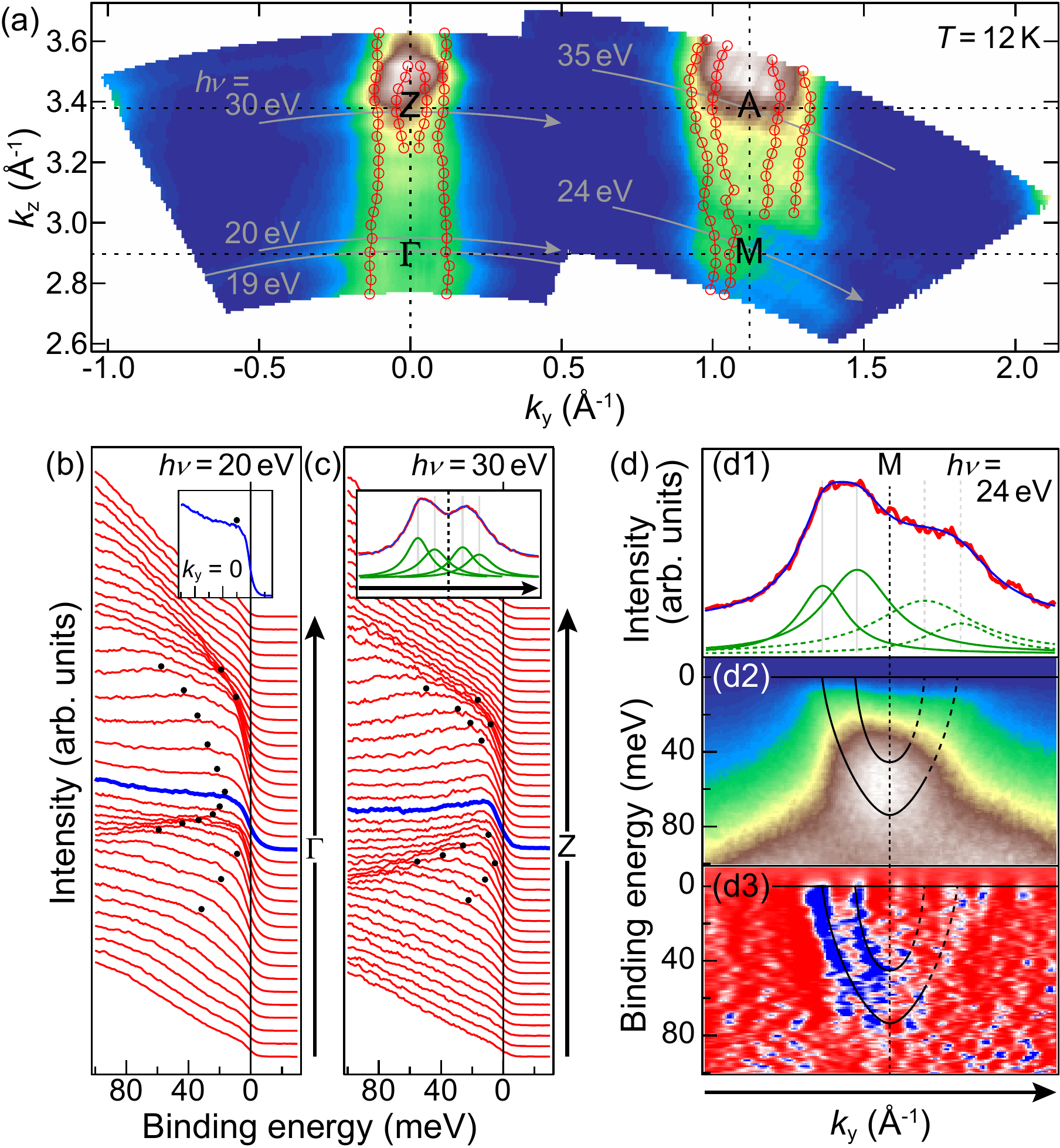}
\caption{
(color online) (a) Out-of-plane FS images.
The open circles denote the FSs determined by the peak position of the MDCs.
See text for details.
(b,c)~ARPES spectra near the $\Gamma$ and Z points.
The insets are the ARPES spectra at $k_y=0$ at the $\Gamma$ point and the MDC fitting at  $E_{\rm F}$ around the Z point.
(d) ARPES results at the M point.
(d1) MDC fitting at the $E_{\rm F}$, (d2) ARPES image and (d3) the MDC's second-derivative image of (d2).
}
\label{fig:one}
\end{center}
\end{figure}
\begin{figure}[b]
\begin{center}
\includegraphics[width=8cm,clip]{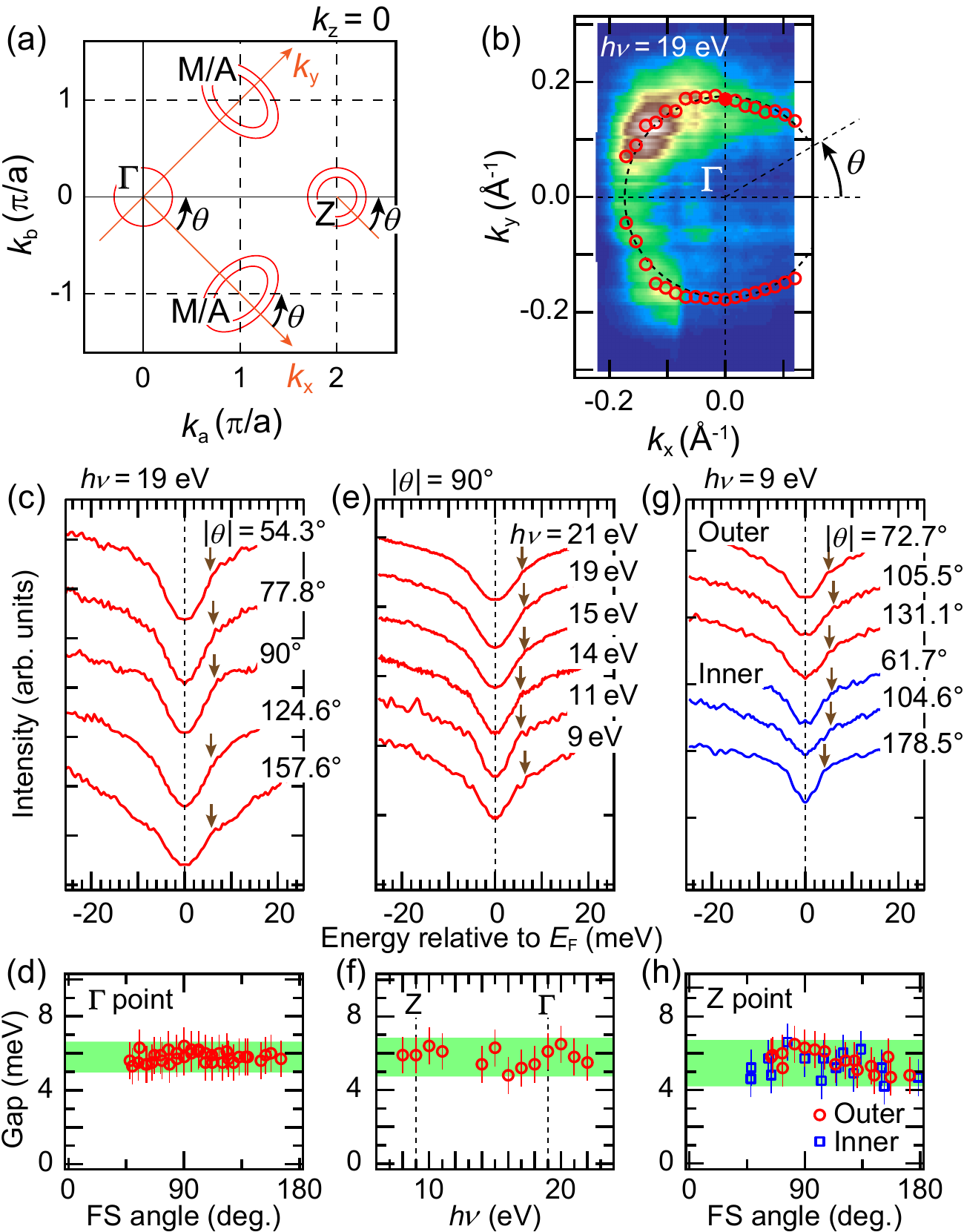}
\caption{
(color online) 3D SC-gap structures at hole FSs.
(a)~Schematic FS image at $k_{z}=0$.
(b)~In-plane FS image at $h\nu=19$~eV around the $\Gamma$~point.
(c)~Symmetrized ARPES spectra with various $\theta$ around the $\Gamma$~point.
(d)~$\theta$ dependence of the SC-gap size around the $\Gamma$~point.
(e)~Symmetrized ARPES spectra with various $h\nu$ at the outer hole FS denoted by the solid circle in panel~(b).
(f)~$h\nu$ dependence ($k_z$ dependence) of the SC-gap size in $h\nu=8$--22~eV in 1 eV steps.
(g)~Symmetrized ARPES spectra of the outer and inner hole FSs at $h\nu=9$~eV around the Z~point.
(h)~$\theta$ dependence of the SC-gap size around the Z~point.
}
\label{fig:two}
\end{center}
\end{figure}

Figure~\ref{fig:one}(a) shows the out-of-plane FSs.
The open circles in Fig.~\ref{fig:one}(a) denote the peak position of the momentum-distribution curves (MDCs) determined by the fitting using a Lorentzian with a constant background, as shown in the inset of Fig.~\ref{fig:one}(c) and in Fig.~\ref{fig:one}(d1).
Hereafter, we use open circles on FS images of each figure in the same manner.
We examine the band character of each FS ($\Gamma$, Z, and M points) from the ARPES data, as shown in Figs.~\ref{fig:one}(b)--\ref{fig:one}(d) along each line cut in Fig.~\ref{fig:one}(a).
The solid circles in Figs.~\ref{fig:one}(b) and~\ref{fig:one}(c) denote the band dispersions determined by the peak and shoulder structure positions of the ARPES spectra; there are two hole bands near $E_{\rm F}$ at both the  $\Gamma$ and Z points.
At the  $\Gamma$ point, one hole band crosses $E_{\rm F}$, while the other has a top at about 20~meV without crossing $E_{\rm F}$, as shown in the inset of Fig.~\ref{fig:one}(b).
At the Z~point, both hole bands cross $E_{\rm F}$, as shown in the inset of Fig.~\ref{fig:one}(c), which is the MDC fitting at $E_{\rm F}$.
At the M point of the zone corner, we confirm the existence of two electron bands with bottoms at about 40 and 80 meV from the MDC fitting at $E_{\rm F}$, from the ARPES image, and from the MDC's second-derivative image, as shown in Figs.~\ref{fig:one}(d1)--\ref{fig:one}(d3), respectively.

Figures~\ref{fig:two}(a) and~\ref{fig:two}(b) show a schematic FS image at $k_z=0$ and the in-plane FS at $h\nu=19$~eV around the $\Gamma$~point, respectively.
Figure~\ref{fig:two}(c) shows the symmetrized ARPES spectra with the selected FS angle ($\theta$) around the $\Gamma$~point.
Each spectrum is normalized to the integrated intensity between 18~and~22~meV.
Hereafter, all the spectra are normalized in the same manner.
Each spectrum has a shoulder structure due to the SC gap at about 5--7~meV, as marked by arrows~\cite{ref-coherence-peak1}, and its $\theta$ dependence is summarized in Fig.~\ref{fig:two}(d).
Figures~\ref{fig:two}(e) and~\ref{fig:two}(f) show the $h\nu$ ($k_z$)-dependent symmetrized ARPES spectra and the SC-gap size at the outer hole FS at $\theta=90^{\circ}$, respectively.
Figures~\ref{fig:two}(g) and~\ref{fig:two}(h) show the  $\theta$-dependent symmetrized ARPES spectra and the SC-gap size of the two (inner and outer) hole FSs around the Z~point, respectively.
Consequently, by combination of the results obtained along the  $\Gamma$-Z line, we found that the SC gaps of the two hole FSs are nearly isotropic without any nodes in 3D momentum space.

\begin{figure*}[t]
\begin{center}
\includegraphics[width=15.5cm,clip]{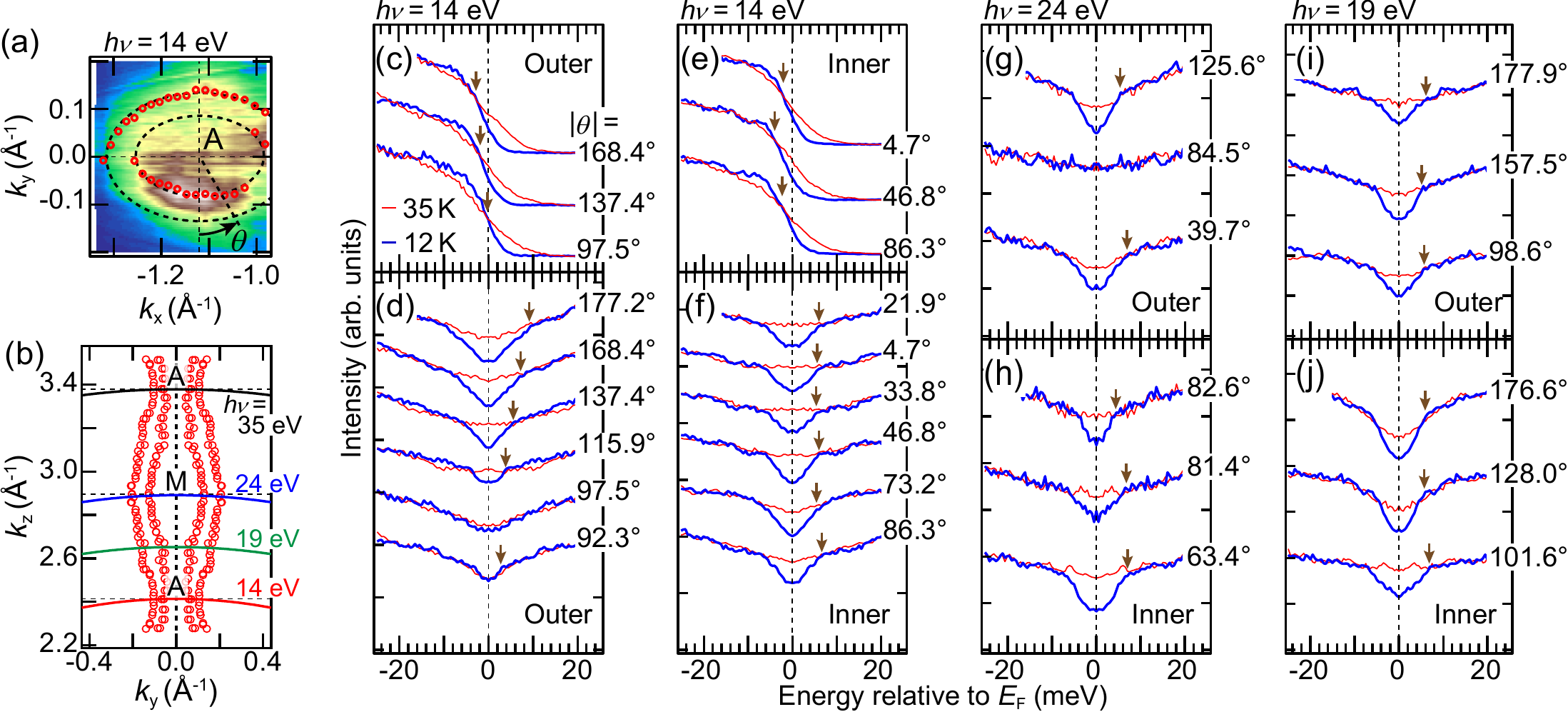}
\caption{
(color online) 3D SC-gap structures at electron FSs.
(a)~In-plane FS image at $h\nu=14$~eV around the A~point.
(b)~$k_z$ position of each SC-gap measurement around the electron FSs as a function of $k_y$.
(c--f)~$T$-dependent and symmetrized ARPES spectra at $T=35$~K ($> T_{c}$) and 12~K ($< T_c$) of the outer~(c, d) and inner~(e, f) electron FSs at $h\nu=14$~eV.
(g--j)~$T$-dependent symmetrized ARPES spectra at the outer~(g, i) and inner~(h, j) electron FSs at $h\nu=24$ and $19$~eV.
}
\label{fig:three}
\end{center}
\end{figure*}

We now turn our attention to the SC-gap structure of the ellipsoid-shaped electron FSs at the zone corner. 
Figure~\ref{fig:three}(a) shows the in-plane FSs at $h\nu=14$~eV around the A~point [Fig.~\ref{fig:three}(b)].
There are two FSs similar to those in Fig.~\ref{fig:one}, but the matrix element is different.
This is because the experimental configuration and photon energy used are different from those used in Fig.~\ref{fig:one}, which were measured at $k_x=0$ with $h\nu=20$--40~eV.
The temperature ($T$)-dependent ARPES spectra at $35$~K ($> T_c$) and $12$~K ($< T_c$) at the outer electron FS are shown in Fig.~\ref{fig:three}(c).
At $|\theta|=168.4^{\circ}$ on the shorter axis, the crossing point of the $T$-dependent spectra is located in the occupied states, thus suggesting the opening of the SC gap.
At $|\theta|=137.4^{\circ}$, the crossing point shifts toward $E_{\rm F}$, and then, at $|\theta|=97.5^{\circ}$ on the longer axis, the crossing point is exactly at $E_{\rm F}$, thus suggesting the closing of the SC gap.
The $T$-dependent symmetrized ARPES spectra with various $\theta$ of the outer electron FS are shown in Fig.~\ref{fig:three}(d).
Around the shorter axis ($|\theta|=180^{\circ}$), the spectra at $T=12$~K ($<T_c$) have shoulder structures marked by arrows, indicating the opening of the SC gap.
Toward the longer axis ($|\theta|=90^{\circ}$), the energy of the shoulder decreases and finally vanishes at $|\theta|=97.5^{\circ}$.
The symmetrized ARPES spectrum at  $|\theta|=92.3^{\circ}$ shows a small dip at about 3~meV, which may be caused by loop-like nodes, as will be discussed later.
In the same manner, we show the $T$-dependent and symmetrized ARPES spectra at the inner electron FS in Figs.~\ref{fig:three}(e) and~\ref{fig:three}(f), respectively.
In contrast with those of the outer electron FS, the crossing point and SC-gap size of the inner FS are almost unchanged.
A similar $\theta$ dependence of the SC gap is also observed at $h\nu=24$~eV around the M~point, as shown in Figs.~\ref{fig:three}(g) and~\ref{fig:three}(h).

For a more detailed determination of the 3D SC-gap structure in the electron FSs, we investigate the electron FSs at the middle of the M and A~points taken at $h\nu=19$~eV, as shown in Fig.~\ref{fig:three}(b).
In the same manner as that for the A~point, the $T$-dependent symmetrized ARPES spectra of the outer and inner electron FSs are shown in Figs.~\ref{fig:three}(i)~and~\ref{fig:three}(j), respectively.
In contrast with the A~and M~points, the SC-gap size of both electron FSs are almost constant.

\begin{figure}[b]
\begin{center}
\includegraphics[width=8cm,clip]{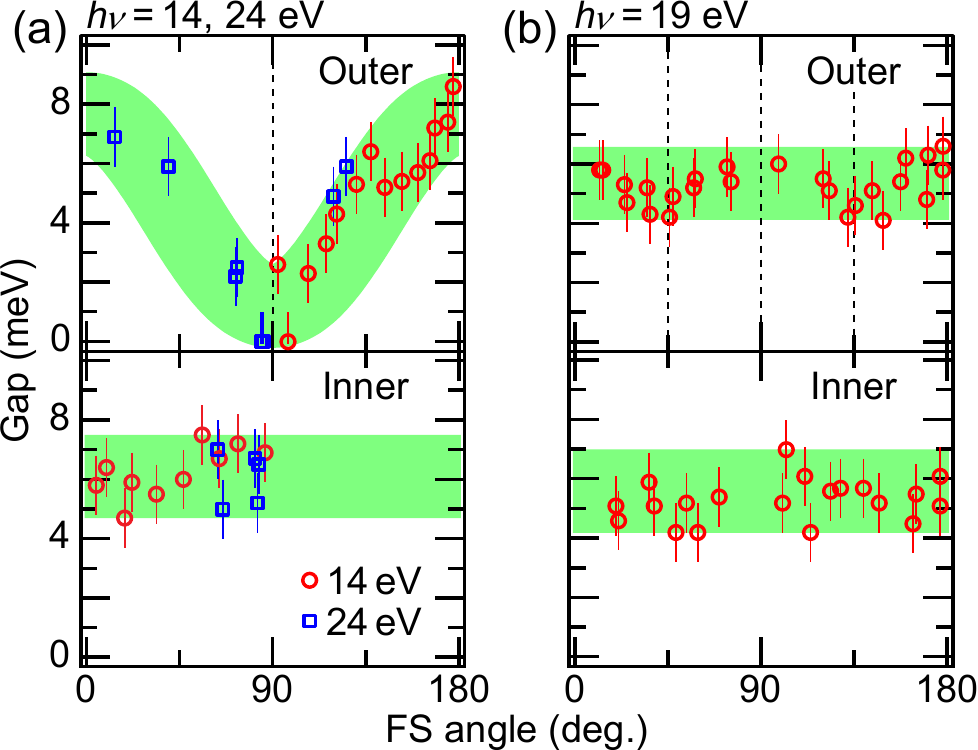}
\caption{
(color online) $\theta$ dependences of the SC-gap sizes of the outer and inner electron FSs at $h\nu=14$ and $24$~eV~(a) and at $h\nu=19$~eV~(b).
}
\label{fig:four}
\end{center}
\end{figure}

We summarize the $\theta$ dependences of the SC-gap sizes of the outer and inner electron FSs around both the A~($h\nu=14$~eV) and M~($h\nu=24$~eV)~points and around the middle of the M and A~points~($h\nu=19$~eV) in Figs.~\ref{fig:four}(a)~and~\ref{fig:four}(b), respectively.
The SC-gap size of the outer electron FS around the A and M points has a maximum of about 8~meV along the shorter axis ($|\theta|=0, 180^{\circ}$).
Toward the longer axis ($|\theta|=90^{\circ}$), the SC-gap size becomes smaller and a node or a gap minimum exists.
On the other hand, at the inner electron FS, the SC gap is nearly isotropic at about 5--7~meV.
Around the middle of the M and A~points~($h\nu=19$~eV); however, both electron FSs have an SC gap at 4--7~meV without nodes, while the outer electron FS has a possible fourfold symmetry within experimental resolution.
Our observations therefore suggest that nodes or gap minima exist only around the longer axis of the ellipsoid-shaped outer electron FS around the M~and A~points.
The reproducibility of the SC gap around the hole FSs was checked by repeated experiments.
For the electron FSs, it was checked by comparing the spectra of the sample cleaved at $T=12$~K with those at $T=12$~K after cleaving at $T=35$~K.
Moreover, it was confirmed from the similarity of the $\theta$ dependences of the SC gaps at the A~($h\nu=14$~eV) and M~($h\nu=24$~eV)~points.

On the basis of the obtained 3D SC-gap structure, we now discuss the mechanism of the appearance of the superconductivity of this system.
In the Ba(Fe$_{1-x}$Co$_x$)$_2$As$_2$ system, both spin fluctuation~\cite{ref16} and the antiferromagnetic scattering vector  {\boldmath $Q$}$_{\rm AF}$ between the hole and electron FSs~\cite{ref17} in the SC region have been reported.
In the heavily overdoped non-SC region, there is no interband scattering owing to the absence of a hole FS~\cite{ref_sekiba}.
These results imply that interband scattering between the hole and electron FSs is important for the appearance of superconductivity; that is, interband scattering seems to be more important for the appearance of superconductivity.
Moreover, the softening of the $C_{66}$ mode~\cite{ref20} and the orbital-polarized electronic structure~\cite{ref21} also suggest the existence of orbital fluctuation.
Hence, not only spin fluctuation but also orbital fluctuation might be involved in the appearance of superconductivity in this system.

According to the theory of the 3D SC gap of the BaFe$_2$(As$_{1-x}$P$_x$)$_2$ system, if a $d_{xy}$ hole FS exists, the fully opened anisotropic SC gap on both electron FSs should appear owing to the spin fluctuation~\cite{ref15}.
Actually, our recent polarization-dependent ARPES study of the Ba(Fe$_{1-x}$Co$_x$)$_2$As$_2$ system suggests the existence of a $d_{xy}$ hole FS~\cite{ref_hajiri}.
On the other hand, owing to orbital fluctuation, both electron FSs have a nearly isotropic SC gap~\cite{ref15}.
If spin and orbital fluctuations compete, both electron FSs have an anisotropic nodal SC gap (with loop-like nodes) depending on the degree of competition~\cite{ref15}.
Our present ARPES results of the Ba(Fe$_{1-x}$Co$_x$)$_2$As$_2$ system suggest different anisotropies between the outer and inner electron FSs around the M and A points, while the above theories predict the same anisotropy between both electron FSs.
Since the different anisotropes obtained are considered to be related to the appearance of superconductivity, further theoretical and also experimental studies are needed.

In summary, we performed 3D-ARPES measurements of slightly overdoped Ba(Fe$_{1-x}$Co$_x$)$_2$As$_2$ ($x=0.1$) to clarify the 3D SC-gap structure.
We observed that two hole FSs and an inner electron FS have a nearly isotropic SC gap, and that the outer electron FS has an anisotropic SC gap with nodes or gap minima.
These results provide a new insight into the appearance of the superconductivity of iron pnictide superconductors.

\begin{acknowledgment}


The authors gratefully acknowledge M. Sakai for his technical assistance during the experiments.
Part of this work was supported by the Use-of-UVSOR Facility Program (BL7U, 2012) of the Institute for Molecular Science.
The work at DGIST was partially supported by the Leading Foreign Research Institute Recruitment Program (Grant No. 2012K1A4A3053565) and Basic Science Research Program (NRF-2013R1A1A2009778) through the National Research Foundation of Korea funded by MEST. 
T.H. was supported by a Grant-in-Aid for JSPS Fellows.

\end{acknowledgment}



\begin{thebibliography}{99}
 \bibitem{ref1}Y. Kamihara, T. Watanabe, M. Hirano, and H. Hosono, J. Am. Chem. Soc. {\bf130}, 3296 (2008).
 \bibitem{ref2}K. Kuroki, H. Usui, S. Onari, R. Arita, and H. Aoki, Phys. Rev. B {\bf79}, 224511 (2009).
 \bibitem{ref3}H. Kontani and S. Onari, Phys. Rev. Lett. {\bf104}, 157001 (2010).
 \bibitem{ref4}K. Hashimoto, T. Shibauchi, S. Kasahara, K. Ikada, S. Tonegawa, T. Kato, R. Okazaki, C. J. van der Beek, M. Konczykowski, H. Takeya, K. Hirata, T. Terashima, and Y. Matsuda, Phys. Rev. Lett. {\bf102}, 207001 (2009).
 \bibitem{ref5}J. K. Dong, S. Y. Zhou, T. Y. Guan, H. Zhang, Y. F. Dai, X. Qiu, X. F. Wang, Y. He, X. H. Chen, and S. Y. Li, Phys. Rev. Lett. {\bf104}, 087005 (2010).
 \bibitem{ref6}H. Ding, P. Richard, K. Nakayama, T. Sugawara, T. Arakane, Y. Sekiba, A. Takayama, S. Souma, T. Sato, T. Takahashi, Z. Wang, X. Dai, Z. Fang, G. F. Chen, J. L. Luo, and N. L. Wang, Europhys. Lett. {\bf83}, 47001 (2008).
 \bibitem{ref7}Y.-M. Xu, Y.-B. Huang, X.-Y. Cui, E. Razzoli, M. Radovic, M. Shi, G.-F. Chen, P. Zheng, N.-L. Wang, C.-L. Zhang, P.-C. Dai, J.-P. Hu, Z. Wang, and H. Ding, Nat. Phys. {\bf7}, 198 (2011).
 \bibitem{ref8}K. Okazaki, Y. Ota, Y. Kotani, W. Malaeb, Y. Ishida, T. Shimojima, T. Kiss, S. Watanabe, C.- T. Chen, K. Kihou, C. H. Lee, A. Iyo, H. Eisaki, T. Saito, H. Fukazawa, Y. Kohori, K. Hashimoto, T. Shibauchi, Y. Matsuda, H. Ikeda, H. Miyahara, R. Arita, A. Chainani, and S. Shin, Science {\bf337}, 1314 (2012).
 \bibitem{ref9}W. Malaeb, T. Shimojima, Y. Ishida, K. Okazaki, Y. Ota, K. Ohgushi, K. Kihou, T. Saito, C. H. Lee, S. Ishida, M. Nakajima, S. Uchida, H. Fukazawa, Y. Kohori, A. Iyo, H. Eisaki, C.-T. Chen, S. Watanabe, H. Ikeda, and S. Shin, Phys. Rev. B {\bf86}, 165117 (2012).
 \bibitem{ref10}K. Hashimoto, M. Yamashita, S. Kasahara, Y. Senshu, N. Nakata, S. Tonegawa, K. Ikada, A. Serafin, A. Carrington, T. Terashima, H. Ikeda, T. Shibauchi, and Y. Matsuda, Phys. Rev. B {\bf81}, 220501(R) (2010).
 \bibitem{ref11}Y. Nakai, T. Iye, S. Kitagawa, K. Ishida, S. Kasahara, T. Shibauchi, Y. Matsuda, and T. Terashima, Phys. Rev. B {\bf81}, 020503(R) (2010).
 \bibitem{ref12}Y. Zhang, Z. R. Ye, Q. Q. Ge, F. Chen, J. Jiang, M. Xu, B. P. Xie, and D. L. Feng, Nat. Phys. {\bf8}, 371 (2012).
 \bibitem{ref13}T. Yoshida, S. Ideta, T. Shimojima, W. Malaeb, K. Shinada, H. Suzuki, I. Nishi, A. Fujimori, K. Ishizaka, S. Shin, Y. Nakashima, H. Anzai, M. Arita, A. Ino, H. Namatame, M. Taniguchi, H. Kumigashira, K. Ono, S. Kasahara, T. Shibauchi, T. Terashima, Y. Matsuda, M. Nakajima, S. Uchida, Y. Tomioka, T. Ito, K. Kihou, C. H. Lee, A. Iyo, H. Eisaki, H. Ikeda, R. Arita, T. Saito, S. Onari, and H. Kontani, arXiv: 1301.4818.
 \bibitem{ref14}K. Suzuki, H. Usui, and K. Kuroki, J. Phys. Soc. Jpn. {\bf80}, 013710 (2011).
 \bibitem{ref15}T. Saito, S. Onari, and H. Kontani, Phys. Rev. B {\bf88}, 045115 (2013).
 \bibitem{ref16}F. L. Ning, K. Ahilan, T. Imai, A. S. Sefat, M. A. McGuire, B. C. Sales, D. Mandrus, P. Cheng, B. Shen, and H.-H. Wen, Phys. Rev. Lett. {\bf104}, 037001 (2010).
 \bibitem{ref17}M. Lumsden, A. Christianson, D. Parshall, M. Stone, S. Nagler, G. MacDougall, H. Mook, K. Lokshin, T. Egami, D. L. Abernathy, E. Goremychkin, R. Osborn, M. McGuire, A. Sefat, R. Jin, B. C. Sales, and D. Mandrus, Phys. Rev. Lett. {\bf102}, 107005 (2009).
 \bibitem{ref18}K. Kuroki, S. Onari, R. Arita, H. Usui, Y. Tanaka, H. Kontani, and H. Aoki, Phys. Rev. Lett. {\bf101}, 087004 (2008).
 \bibitem{ref19}I. I. Mazin, D. J. Singh, M. D. Johannes, and M. H. Du, Phys. Rev. Lett. {\bf101} (2008) 057003. 
 \bibitem{ref20}M. Yoshizawa, R .Kamiya, R. Onodera, Y. Nakanishi, K. Kihou, H. Eisaki, and C. H. Lee, J. Phys. Soc. Jpn. {\bf81}, 024604 (2012).
 \bibitem{ref21}M. Yi, D. Lu, J.-H. Chu, J. G. Analytis, A. P. Sorini, A. F. Kemper, B. Moritz, S.-K. Mo, R. G. Moore, M. Hashimoto, W.-S. Lee, Z. Hussain, T. P. Devereaux, I. R. Fisher, and Z.-X. Shen, Proc. Natl. Acad. Sci. {\bf108}, 6878 (2011).
 \bibitem{ref22}L. Luan, O. M. Auslaender, T. M. Lippman, C. W. Hicks, B. Kalisky, J.-H. Chu, J. G. Analytis, I. R. Fisher, J. R. Kirtley, and K. A. Moler, Phys. Rev. B {\bf81}, 100501(R) (2010).
 \bibitem{ref23}M. A. Tanatar, J.-Ph. Reid, H. Shakeripour, X. G. Luo, N. Doiron-Leyraud, N. Ni, S. L. Bud'ko, P. C. Canfield, R. Prozorov, and L. Taillefer, Phys. Rev. Lett. {\bf104}, 067002 (2010).
\bibitem{ref24}J.-Ph. Reid, M. A. Tanatar, X. G. Luo, H. Shakeripour, N. Doiron-Leyraud, N. Ni, S. L. Bud'ko, P. C. Canfield, R. Prozorov, and L. Taillefer, Phys. Rev. B {\bf82}, 064501 (2010).
\bibitem{ref25}J. S. Kim, B. D. Faeth, Y. Wang, P. J. Hirschfeld, G. R. Stewart, K. Gofryk, F. Ronning, A. S. Sefat, K. Y. Choi, and K. H. Kim, Phys. Rev. B {\bf86}, 014513 (2012).
\bibitem{ref26}D.-J. Jang, A. B. Vorontsov, I. Vekhter, K. Gofryk, Z. Yang, S. Ju, J. B. Hong, J. H. Han, Y. S. Kwon, F. Ronning, J. D. Thompson, and T. Park, New J. Phys. {\bf13}, 023036 (2011).
\bibitem{ref27}K. Terashima, Y. Sekiba, J. H. Bowen, K. Nakayama, T. Kawahara, T. Sato, P. Richard, Y.-M. Xu, L. J. Li, G. H. Cao, Z.-A. Xu, H. Ding, and T. Takahashi: Proc. Natl. Acad. Sci. {\bf106}, 7330 (2009).
\bibitem{ref28}S. Kimura, T. Ito, M. Sakai, E. Nakamura, N. Kondo, T. Horigome, K. Hayashi, M. Hosaka, M. Katoh, T. Goto, T. Ejima, and K. Soda, Rev. Sci. Instrum. {\bf81}, 053104 (2010).
\bibitem{ref-coherence-peak1} The coherence peak seems to be weak. This may originate from an impurity scattering effect in unconventional superconductors. See also Y. Yin, M. Zech, T. L. Williams, X. F. Wang, G. Wu, X. H. Chen, and J. E. Hoffman, Phys. Rev. Lett. {\bf102}, 097002 (2009); and G. Preosti, H. Kim, and P. Muzikar, Phys. Rev. B {\bf50}, 1259 (1994).
\bibitem{ref_sekiba}Y. Sekiba, T. Sato, K Nakayama, K. Terashima, P. Richard, J. H. Bowen, H. Ding, Y.-M. Xu, L. J. Li, G. H. Cao, Z.-A. Xu, and T. Takahashi, New J. Phys. {\bf11}, 025020 (2009).
\bibitem{ref_hajiri}T. Hajiri, T. Ito, M. Matsunami, B. H. Min, Y. S. Kwon, and S. Kimura, JPS Conf. Proc. {\bf3}, 015028 (2014).


\end{thebibliography}
\end{document}